\documentclass[final]{aipproc}

\layoutstyle{6x9}

\usepackage{amsmath}

\DeclareMathOperator{\tr}{Tr}

\begin{document}

\title{Sivers Effect Asymmetries in Hadronic Collisions}


\author{C. J. Bomhof}{
address={Department of Physics and Astronomy, Vrije Universiteit Amsterdam,\\
NL-1081 HV Amsterdam, the Netherlands.}
}

\author{P. J. Mulders}{
address={Department of Physics and Astronomy, Vrije Universiteit Amsterdam,\\
NL-1081 HV Amsterdam, the Netherlands.}
}

\begin{abstract}
We argue that weighted azimuthal single spin asymmetries in back-to-back jet or pion production in polarized proton-proton scattering can be written as convolutions of universal distribution and fragmentation functions with gluonic pole cross sections as hard functions.
Gluonic pole cross sections are gauge-invariant weighted sums of Feynman diagrams.
The weight factors are a direct consequence of the (diagram-dependence of) gauge links.
The best known consequence of the gauge links is the generation of the Sivers effect that is a source for single-spin asymmetries.
Moreover, due to the dependence of the gauge links on the color-flow of the hard diagram the Sivers effect in SIDIS enters with opposite sign as it does in Drell-Yan scattering.
The weight factors in the gluonic pole cross sections are the appropriate generalizations to more complicated processes of this relative sign difference.
Furthermore, it is argued that the gluon-Sivers effect appears in twofold.
\end{abstract}

\maketitle

\begin{figure}
\includegraphics[width=4cm]{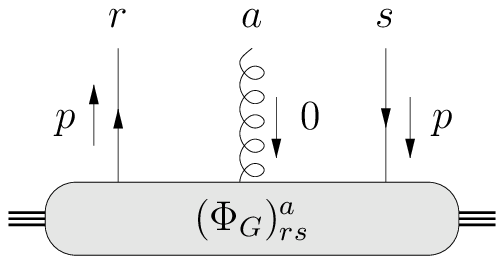}\hspace{1cm}
\includegraphics[width=4cm]{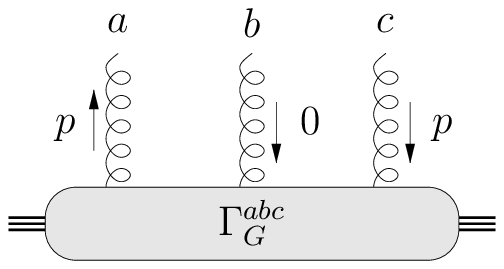}
\caption{Different types of gluonic pole matrix elements (GPME's):
the quark-GPME $\Phi_G{=}(\Phi_G)_{rs}^a(t^a)_{sr}{=}\tr[\Phi_G^at^a]$
and the two gluon-GPME's 
$\Gamma_G^{(f)}{=}\Gamma_G^{abc}(t^a)_{cb}{=}\Gamma_G^{abc}if^{abc}$ and $\Gamma_G^{(d)}{=}\Gamma_G^{abc}d^{abc}$.
\label{GLUONgluonicPOLES}}
\end{figure}

For (semi)-inclusive measurements the cross section in hard scattering processes factorizes into a product of a hard squared amplitude and parton distribution and fragmentation functions.
The $k_T$-dependent distribution functions can be regarded as projections of the transverse momentum dependent (TMD) correlator
\begin{equation}\label{TMDcorrelator}
\Phi^{[\mathcal U]}(x{,}k_T)
={\int}\frac{\mathrm d(\xi{\cdot}P)\mathrm d^2\xi_T}{(2\pi)^3}\ e^{ik\cdot\xi}
\langle P{,}S|\,\overline\psi(0)\,\mathcal U\!(0{,}\xi)\,
\psi(\xi)\,|P{,}S\rangle\big\rfloor_{\text{light-front}}\ ,
\end{equation}
The gauge link 
$\mathcal U\!(0{,}\xi)
{=}\mathcal P\exp[-ig\int_C\mathrm dz^\mu\,A_\mu^a(x)t^a]$ 
is a path-ordered exponential.
Its presence in the hadronic matrix element is required by gauge-invariance.
In the TMD correlator~\eqref{TMDcorrelator} the integration path $C$ in the gauge link is process-dependent.
In the diagrammatic approach it arises by resumming all collinear gluon interactions between the soft and the hard part~\cite{Boer:1999si,Belitsky:2002sm,Boer:2003cm}.
Consequently, the integration path $C$ is not a freedom of choice,
but it is fixed by the color-flow of the hard part of the scattering process~\cite{Bomhof:2006dp}.
A well known example is semi-inclusive deep-inelastic scattering (SIDIS)
where the resummation of all collinear gluon interactions leads to a future pointing Wilson line.
Another example is Drell-Yan scattering (DY) where it gives a past pointing Wilson line.
In the $k_T$-integrated correlator 
$\Phi(x){=}{\int}\mathrm d^2k_T\,\Phi^{[\mathcal U]}(x{,}k_T)$ 
all process-dependence of the gauge link has been integrated away,
leaving just a straight Wilson line in the light-cone $n$-direction
(the direction conjugate to $P$ and perpendicular to $k_T$).
However, in the transverse momentum weighted correlator (transverse moment)
\begin{equation}\label{TransverseMoment}
\Phi_\partial^{[\mathcal U]\alpha}(x)
={\int}\mathrm d^2k_T\;k_T^\alpha\;\Phi^{[\mathcal U]}(x{,}k_T)
=\widetilde\Phi{}_\partial^\alpha(x)
+C_G^{[\mathcal U]}\,\pi\Phi_G^\alpha(x{,}x)\ ,
\end{equation}
there is still a (sub)process-dependence residing in the multiplicative factor 
$C_G^{[\mathcal U]}$ that is completely determined by the gauge link and, hence, by (the color-flow of) the hard squared amplitude.
The matrix elements in the second equality of expression~\eqref{TransverseMoment} are defined in~\cite{Boer:2003cm}.
Just as the $k_T$-integrated correlator they only contain
straight Wilson lines in the light-cone $n$-direction.
The correlator $\widetilde\Phi{}_\partial^\alpha(x)$ is even under time-reversal,
while $\Phi_G^\alpha(x{,}x)$ is $T$-odd.
The latter matrix element is called a gluonic pole matrix element,
since it is a matrix element of two quark fields and a zero-momentum gluon
(see Figure~\ref{GLUONgluonicPOLES}).
It is proportional to the Qiu-Sterman matrix element $T_F(x{,}x)$
and the (first moment of) the Sivers function $f_{1T}^{\perp(1)}(x)$ is contained in its parametrization~\cite{Boer:2003cm}.
From~\eqref{TransverseMoment} it is seen that as a consequence of the gauge link the gluonic pole matrix element is mul\-tiplied by a (sub)process dependent factor $C_G^{[\mathcal U]}$.
Since this factor multiplies the gluonic pole matrix element, it is called a gluonic pole strength.
The future pointing Wilson line in SIDIS leads to Eq.~\eqref{TransverseMoment} with $C_G^{\text{SIDIS}}{=}{+}1$,
while the past pointing Wilson line in DY gives the gluonic pole strength 
$C_G^{\text{DY}}{=}{-}1$.
From this observation follows the important conclusion that the Sivers effect appears with opposite signs in SIDIS and DY~\cite{Brodsky:2002rv,Collins:2002kn}:
\begin{subequations}\label{SiversEffect}
\begin{alignat}{3}
&\text{Sivers effect in SIDIS:}&\qquad
&\mathrm d\sigma_{\ell H{\rightarrow}\ell h X}&&\sim \ 
+f_{1T}^{\perp(1)}(x)
\,\mathrm d\hat\sigma_{\ell q{\rightarrow}\ell q}\,D_1(z)\ ,\\
&\text{Sivers effect in DY:}&
&\mathrm d\sigma_{HH^\prime{\rightarrow}\ell\bar\ell X}&\ &\sim \ 
-f_{1T}^{\perp(1)}(x)\,\bar f_1(x')
\,\mathrm d\hat\sigma_{q\bar q{\rightarrow}\ell\bar\ell}\ .
\end{alignat}
\end{subequations}

\begin{figure}
\begin{minipage}[t]{0.32\textwidth}
\centering
\includegraphics[width=\textwidth]{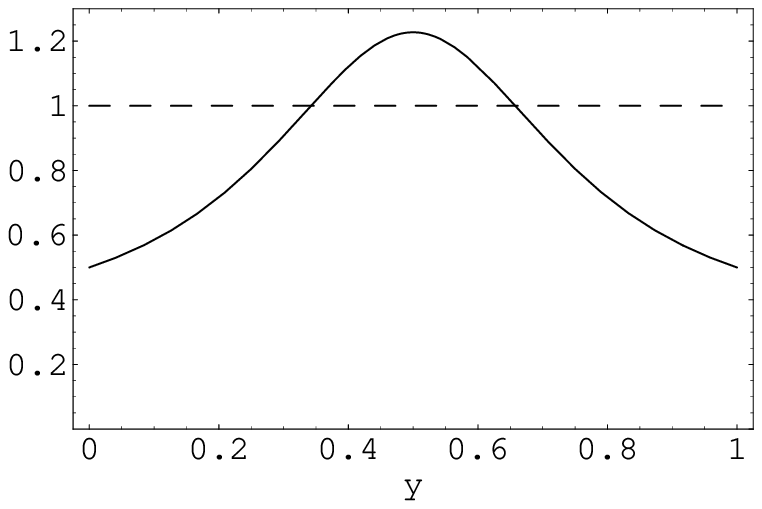}
\hspace*{6mm}${\scriptstyle\mathrm d\hat\sigma_{[q]q\rightarrow qq}/
\mathrm d\hat\sigma_{qq\rightarrow qq}}$
\end{minipage}
\begin{minipage}[t]{0.32\textwidth}
\centering
\includegraphics[width=\textwidth]{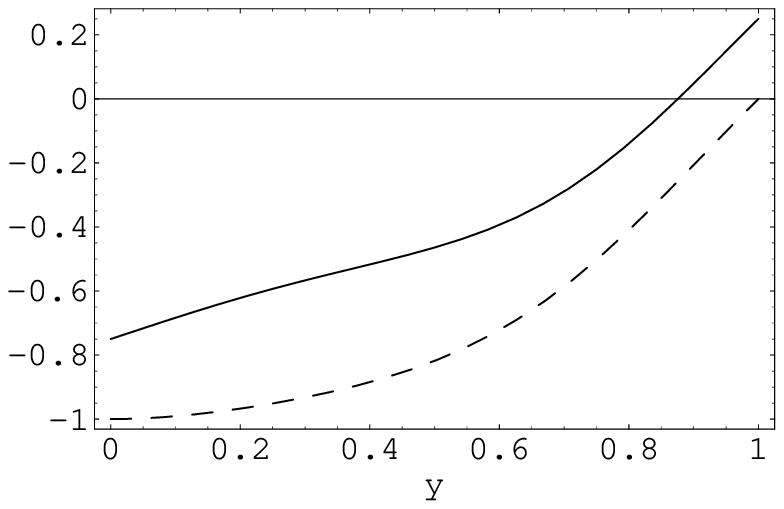}
\hspace*{6mm}${\scriptstyle\mathrm d\hat\sigma_{[q]\bar q\rightarrow q\bar q}/
\mathrm d\hat\sigma_{q\bar q\rightarrow q\bar q}}$
\end{minipage}
\begin{minipage}[t]{0.32\textwidth}
\centering
\includegraphics[width=\textwidth]{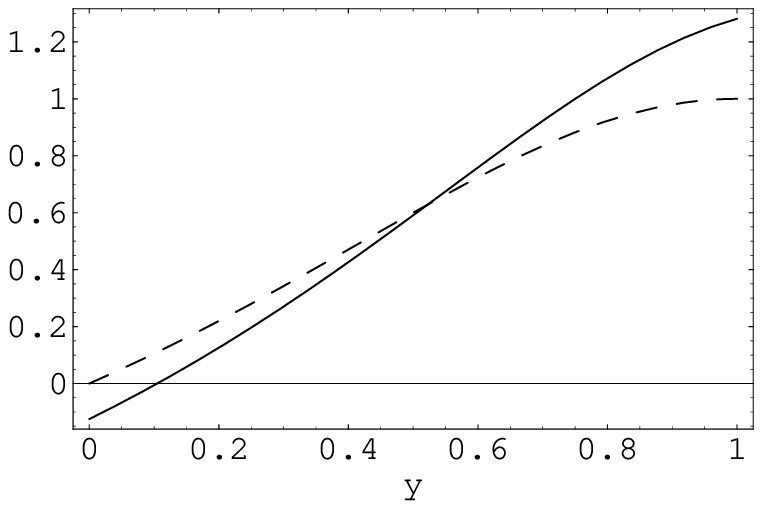}
\hspace*{6mm}${\scriptstyle\mathrm d\hat\sigma_{[q]g\rightarrow qg}/
\mathrm d\hat\sigma_{qg\rightarrow qg}}$
\end{minipage}
\caption{Some examples of ratios of gluonic pole cross sections and partonic cross sections as functions of the variable $y{\equiv}{-}\hat t/\hat s$
for $N{=}3$ (solid line) and $N{\rightarrow}\infty$ 
(dashed line).\label{comparison}}
\end{figure}

Also in processes with more complicated hard functions than those encountered in SIDIS and DY a decomposition as in~\eqref{TransverseMoment} can be made.
However, in those cases the gluonic pole strength is not limited to just a sign~\cite{Bomhof:2004aw,Bacchetta:2005rm,Bomhof:2006ra}.
It becomes particularly interesting when one considers a process where there are different Feynman diagrams that contribute to the hard process.
Since the gauge link depends on the color-flow of the hard diagram
the gluonic pole strength can in general be different for each contribution.
For example, the gluonic pole strengths of the direct scattering contributions to identical quark scattering are $C_G{=}(N^2{-}5)/(N^2{-}1){=}\frac{1}{2}$,
whereas those of the interference diagrams are $C_G{=}{-}(N^2{+}3)/(N^2{-}1){=}{-}\frac{3}{2}$.
Therefore, the Sivers-effect contribution of identical quark scattering is seen to contain the gluonic pole matrix element $f_{1T}^{\perp(1)}(x)$ in the combination
\begin{equation}
f_{1T}^{\perp(1)}(x)\,\bigg(
\big\{{\tfrac{1}{2}}\big\}
\parbox{1.5cm}{\includegraphics[width=1.5cm]{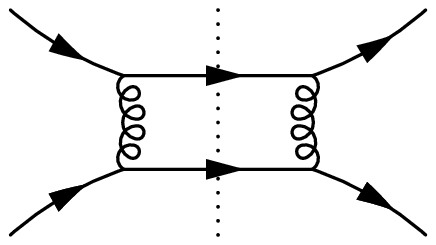}}
+\big\{{\tfrac{1}{2}}\big\}
\parbox{1.5cm}{\includegraphics[width=1.5cm]{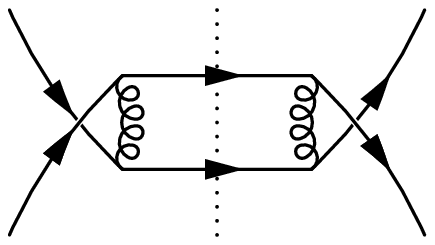}}
-\big\{{-}{\tfrac{3}{2}}\big\}
\parbox{1.5cm}{\includegraphics[width=1.5cm]{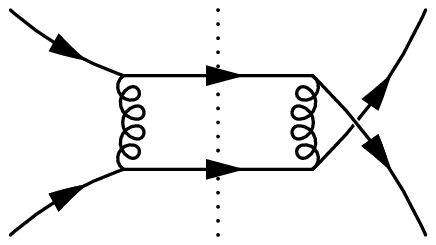}}
-\big\{{-}{\tfrac{3}{2}}\big\}
\parbox{1.5cm}{\includegraphics[width=1.5cm]{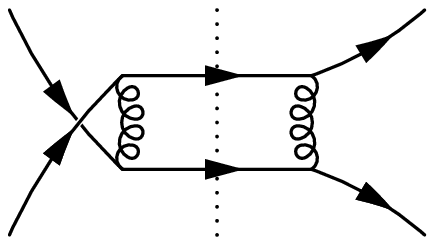}}\;\bigg)\ ,
\end{equation}
rather than with the partonic cross section,
which does not contain the weight factors between braces $\{\cdot\}$.
This observation generalizes to all partonic processes and hence the quark-Sivers function $f_{1T}^{\perp(1)}$ is seen to appear with the hard functions
\begin{equation}\label{Hard}
\mathrm d\hat\sigma_{[q]a\rightarrow bc}
=\sum\nolimits_{D}\;C_G^{[\mathcal U\!(D)]}\,
\mathrm d\hat\sigma_{qa\rightarrow bc}^{[D]}\ ,
\end{equation}
where $\mathrm d\hat\sigma^{[D]}$ is the squared amplitude expression of the Feynman diagram $D$ and the summation runs over all diagrams $D$ that can contribute to the process $qa{\rightarrow}bc$.
The hard functions~\eqref{Hard} are the \emph{gluonic pole cross sections}.
Hence, the gluonic pole cross sections are gauge-invariant weighted sums of Feynman diagrams that are, in general,
distinct from the partonic cross sections that enter in spin-averaged processes
(see Figure~\ref{comparison}).
The bracketed subscript $[q]$ indicates that in this example it is quark $q$ that contributes the gluonic pole.
There seems to be a close relation~\cite{Bomhof:2006ra} 
between these gluonic pole cross sections and the hard functions calculated in~\cite{Qiu:1998ia,Kouvaris:2006zy}.
The underlying reason for this relation needs further investigation.

Using expression~\eqref{TransverseMoment} we have seen that the Sivers effects corresponding to different Feynman diagrams are all proportional to a universal gluonic pole matrix element $f_{1T}^{\perp(1)}(x)$,
with the gluonic pole strengths as proportionality factors.
It has been shown that the relative sign difference for the Sivers effect in SIDIS and DY also holds for the $k_T$-dependent Sivers functions
$(f_{1T}^{\perp})^{\text{DY}}(x{,}k_T)
{=}{-}(f_{1T}^{\perp})^{\text{SIDIS}}(x{,}k_T)$~\cite{Collins:2002kn}.
The underlying reason is that SIDIS and DY involve future and past-pointing Wilson lines, respectively,
that are related by time-reversal.
Going beyond SIDIS and DY,
it seems that one in general has a `jungle' of Wilson lines.
Therefore, it is not clear (at the time of writing) if the proportionality of the first moments of the Sivers functions also extends to the $k_T$-dependent Sivers functions in processes other than SIDIS and DY.
That is the reason why in~\cite{Bacchetta:2005rm,Bomhof:2006ra} 
weighted single-spin asymmetries were considered for back-to-back jet or pion production in proton-proton scattering
($p^\uparrow p{\rightarrow}JJX$, $p^\uparrow p{\rightarrow}\pi\pi X$).
In those processes one can construct asymmetries that at tree-level only contain $k_T$-integrated and $k_T$-weighted correlators with collinear hard functions.
Hence, the Sivers-effect contribution to the azimuthal asymmetry can be expressed as a product of a universal $T$-odd matrix element $f_{1T}^{\perp(1)}(x)$ and universal $T$-even distribution and fragmentation functions corresponding to the other partons by taking the gluonic pole cross sections as hard functions.
As was shown in~\cite{Bomhof:2006ra} 
these conclusions generalize to all the $T$-odd effects described by gluonic pole matrix elements.
In that reference all gluonic pole cross sections that could contribute to azimuthal asymmetries in proton-proton scattering have been calculated.
Whether or not the use of these gluonic pole cross sections rather than the basic partonic cross sections in the factorized form for weighted asymmetries in di-jet production leads to observable effects for the hadronic asymmetry is not clear at present.
This will be a topic for investigation in the near future.
Though the method outlined above remains a conjecture as long as factorization for back-to-back jet (or pion) production has not formally been proved,
we have confidence in the procedure,
since for this process the conclusions hold for the \emph{tree-level} contribution where any soft-factor will (probably) be unity.

Just as in the case of quarks the gluonic pole matrix elements for gluons contain an additional zero-momentum gluon.
However, since these matrix elements contain three gluon fields,
there are two distinct ways of ordering the fields (since the fields appear color-traced in the matrix element).
From a mathematical point of view it is most compelling to take the commutator and anticommutator combinations,
as these correspond to the color-singlet combinations of the colored fields (see Figure~\ref{GLUONgluonicPOLES}).
They involve the antisymmetric $f^{abc}$ and symmetric $d^{abc}$ structure constants of $SU(3)$, respectively.
Hence, there are two distinct gluonic pole matrix elements in the case of gluons and the decomposition of the transverse moment of the TMD gluon correlator
$\Gamma^{[\mathcal {U\!U'}]}\!(x{,}{k_T}\!\!)$ becomes~\cite{Bomhof:2006ra}
\begin{equation}\label{GTM}
\Gamma^{[\mathcal U\mathcal U^\prime]\alpha}_\partial(x)
=\widetilde\Gamma_\partial^\alpha(x)
+C_G^{(f)}\,\pi\Gamma_G^{(f)}{}^{\alpha}(x{,}x)
+C_G^{(d)}\,\pi\Gamma_G^{(d)}{}^{\alpha}(x{,}x)\ .
\end{equation}
Both gluonic pole matrix elements contain a gluon-Sivers-like distribution function in their parametrization.
From the decomposition~\eqref{GTM} it is seen that these two functions will\linebreak[4] appear multiplied by different gluonic pole cross sections,
one containing the weight factors $C_G^{(f)}$ and one containing the weights $C_G^{(d)}$.
Therefore, the gluon-Sivers contribution to the azimuthal asymmetry in di-jet production in proton-proton scattering will have the generic form
\begin{equation}
G_T^{(f)}{}^{(1)}(x)\,
\mathrm d\hat\sigma_{[g]a\rightarrow bc}^{(f)}
\;+\;G_T^{(d)}{}^{(1)}(x)\,
\mathrm d\hat\sigma_{[g]a\rightarrow bc}^{(d)}\ .
\end{equation}
Though the two gluon-Sivers functions will appear multiplied by the same combination of distribution and fragmentation functions,
they are weighted with different gluonic pole cross sections.
This might allow for an experimental disentanglement of the two gluon-Sivers functions.
For example, in direct photon production only the function $G_T^{(d)}{}^{(1)}(x)$ contributes,
while in the gluon-gluon scattering contribution to proton-proton scattering one only has the function $G_T^{(f)}{}^{(1)}(x)$~\cite{Bomhof:2006ra}.
Since the two gluon-Sivers functions are gluonic pole matrix elements containing
(apart from the two gluon fields) an additional zero-momentum gluon,
they have no immediate probabilistic interpretation
(as is also the case for the quark-Sivers function $f_{1T}^{\perp(1)}(x)$).

\vspace*{-2.4mm}

\bibliographystyle{aipproc}
\bibliography{references}

\end{document}